\documentstyle[12pt]{article}
\def \be{\begin{equation}}
\def \ee{\end{equation}}
\def \ba{\begin{array}{l}}
\def \ea{\end{array}}
\def \bq{\begin{eqnarray}}
\def \eq{\end{eqnarray}}
\def \nn{\nonumber\\}
\def \lb{\label}

\def \fr{\frac}

\def \b{\beta}
\def \d{\delta}

\def \z{\zeta}
\def \tl{\tilde}
\def \ol{\overline}

\def \[{\left[}
\def \]{\right]}
\def \({\left(}
\def \){\right)}

\def \J{J_{ij}}

\def \2{\frac{1}{2}}
\def \4{\frac{1}{4}}

\newcommand{\sectio}[1]{\section{#1}\setcounter{equation}{0}}

\begin{document}

\frenchspacing
\setlength{\parskip}{2mm}

\vspace{10mm}

\begin{center}

{\Large \bf Exact solution of the random bipartite matching model}
\vskip .2in
Vik. S. Dotsenko\footnote{On leave from 
Landau Institute for Theoretical Physics, Moscow}
\vskip .1in
Laboratoire de Physique  Theorique des Liquides, \\
UMR 7600, Universite Paris VI, \\
4 place Jussieu, 75252 Paris Cedex 05,
France\\

\end{center}

\vskip .3in

\begin{abstract}
In this paper we present the exact solution for the average
minimum energy of the random bipartite matching model with
an arbitrary finite number of elements where random paired
interactions are described by independent exponential distribution.
This solution confirms the Parisi conjecture proposed 
for this model earlier, as well as the result of the replica solution
of this model in the thermodynamic limit.
\end{abstract}

\newpage

\sectio{The Model}

The model under consideration can be formulated as follows.
We have a society consisting of $N$ "men" 
(labeled by $i = 1,2,...,N$)  and $N$ "women"
(labeled by $j = 1,2,...,N$)
described by a given set of $N^{2}$ random non-negative interactions
$\{\J\}$ between every man and every woman.
The statistics of $\J$'s is defined by a probability distribution
function $P[\J]$.

Then we consider
all possible "marriages" with the strict monogamy: every man can be
connected with one and only one women, and vice versa.
Thus, a particular marriage configuration in this society can be
described by the $N\times N$ permutation matrix $S_{ij}$ with 
the elements taking values 0 or 1 ("0" for all non-coupled pairs of
men and women, and "1" for married couples) constrained by two
conditions:

\be
\lb{1}
\sum_{i=1}^{N} S_{ij} = \sum_{j=1}^{N} S_{ij} = 1
\ee
which allow one and only one "1" in each row and in each column
of the matrix $\hat S$.
The total number of all possible marriage configurations in this
society is thus equal to $N!$.

Now for every marriage configuration $\hat S$ we introduce
the total energy, or total weight (the Hamiltonian):

\be
\lb{2}
H[\hat S;\hat J] = \sum_{i,j=1}^{N} S_{ij} \J
\ee
For a given matrix $\hat S$ this energy is equal to the sum
of $N$ particular $\J$'s (one from each line and each column)
corresponding to the particular
married couples.
In this paper we consider the simplest possible model in which
the interactions $\{\J\}$ are assumed to be independent and 
described by the bounded exponential distribution:

\be
\lb{3}
P[\J] = \prod_{i,j=1}^{N} \exp\( -\J\) \; ; \; \; \; \; \; \;
( 0 \leq \J < +\infty )
\ee
The problem studied below is formulated as follows:
one has to find the value $E_{N}$ of the average (over the distribution
$P[\J]$) {\it minimum} (over all configurations of the permutation
matrix $S_{ij}$) energy (\ref{2}):

\be
\lb{4}
E_{N} = \[\prod_{i,j=1}^{N} \int_{0}^{\infty} d\J \] P[\J]
\min_{S_{ij}} \(\sum_{i,j=1}^{N} S_{ij} \J\)
\ee
Equivalently, in the language of statistical mechanics $E_{N}$ 
can be obtained as the zero-temperature limit of the average free energy:

\bq
\lb{5}
E_{N} &=& -\lim_{\b\to\infty}\fr{1}{\b}
\[\prod_{i,j=1}^{N} \int_{0}^{\infty} d\J \exp\( -\J\)\]
\log\(\sum_{S_{ij}} \exp\[-\b\sum_{i,j=1}^{N} S_{ij} \J\] \)
\nn
\nn
&\equiv & -\lim_{\b\to\infty}\fr{1}{\b}
\ol{\(\log\[\sum_{S_{ij}} \exp\{-\b H[\hat S;\hat J]\}\] \)} 
\eq
Thus, we face the typical problem of statistical mechanics with quenched
disorder: first, for given values of random parameters $\{\J\}$
one has to compute the partition function and the free energy, 
and only after that one has to carry out the averaging over $\J$'s.

In the thermodynamic limit ($N\to\infty$) this problem has been solved
some years ago in the framework of the replica symmetric ansatz
\cite{MP}, yielding the result:

\be
\lb{6}
E_{N\to\infty} = \z(2) = \fr{\pi^{2}}{6}
\ee
In this paper we present the exact solution of this problem for
an arbitrary (finite) value of $N$.

The case $N=1$ is trivial:

\be
\lb{7}
E_{N=1} = 1
\ee
The case $N=2$ is only slightly more complicated, and it can also be
easily calculated explicitly. Here the $2\times 2$ permutation 
matrix $\hat S$ can have only two configurations: 

\be
\lb{8}
\begin{tabular}{|l|l|}
\hline
1 & 0 \\
\hline
0 & 1 \\
\hline
\end{tabular}
\ee
and
\be
\lb{9}
\begin{tabular}{|l|l|}
\hline
0 & 1 \\
\hline
1 & 0 \\
\hline
\end{tabular}
\ee
Thus, according to the definitions (\ref{4}) or (\ref{5}) we
have:

\be
\lb{10}
E_{N=2} = 2 \int_{0}^{\infty}dJ_{11}dJ_{12}dJ_{21}dJ_{22} (J_{11}+J_{22})            
           \exp\{ -J_{11}-J_{12}-J_{21}-J_{22}\} 
           \theta(J_{12}+J_{21}-J_{11}-J_{22})
\ee
Here the $\theta$-function ensures that the state (\ref{8})
has lower energy than the one (\ref{9}) (due to obvious symmetry 
of the system the contribution from the opposite situation turns
out to be the same, and this provides the factor 2 in the above equation).
Simple integration yields:

\be
\lb{11}
E_{N=2} = 1 + \4
\ee
Noting that the result (\ref{6}) for $N=\infty$ 
can be represented also in the form:

\be
\lb{12}
E_{N\to\infty} = \z(2) = \sum_{k=1}^{\infty} \fr{1}{k^{2}}
\ee
and taking into account the results (\ref{7}) and (\ref{11}), 
G.Parisi has recently proposed very elegant conjecture that the 
solution of the problem for arbitrary value of $N$ must be the
following \cite{G}:

\be
\lb{13}
E_{N} = \sum_{k=1}^{N} \fr{1}{k^{2}}
\ee

For the {\it direct} calculation of $E_{N}$ (in the style of eq.(\ref{10})) 
with an arbitrary $N$ one should perform
the integration over the parameters $\{\J\}$
in the constrained positive subspace $\J \geq 0$ of
the $N^{2}$-dimensional space. Since the total number
of states of the $N\times N$ permutation matrix is equal to $N!$ this 
integration is also constrained by $(N! - 1)$ hyperplanes which guarantee
that chosen one particular state has the minimum energy.
One can easily verify that even in the case $N=3$ such calculation 
turns out to be extremely difficult problem.
Nevertheless, simple numerical tests for $N = 3, 4, 5$ proved to be 
compatible with the above conjecture with the precision $\sim 10^{-5}$ \cite{G}.
Moreover, recent analytical studies have provided the exact solution
of this problem up to $N=4$ and the result of this solution confirms
the conjecture (\ref{13}) \cite{BD}. Here we use the original idea 
(proposed by S.Bravyi) of this unpublished work
to prove that the conjecture (\ref{13}) is indeed correct for arbitrary $N$.

\sectio{The Proof}

To ease further presentation of the proof let us introduce the following
notation. The operation of finding the average of minimum energy of the 
$N\times N$ problem 
(defined in eqs.(\ref{4}) or (\ref{5})) will be denoted by the symbol

\be
\lb{14}
{\bf E} \( \; \; \;
\begin{tabular}{|l|l|l|l|l|l|}
\hline
$\;$ & $\;$ & $\;$ & $\;$ & ... & $\;$ \\
\hline
 &  &  &  & ... & \\ 
\hline
 &  &  &  & ... & \\ 
\hline
 &  &  &  & ... & \\ 
\hline
\multicolumn{6}{|c|}{. . . . . . . . . . . . }  \\ 
\hline
 &  &  &  & ... & \\ 
\hline
\end{tabular}
\; \; \; \)  \; \; = \; \; E_{N}
\ee
It is assumed that "empty" boxes in the above matrix actually
contain random elements $\{\J\}$

Let us consider the first line of the random matrix $\J$, and 
among $N$ its elements $J_{1j}$ let us find the minimum one:
$J^{(1)} \equiv \min_{j}(J_{1j})$. Due to obvious symmetry of the 
problem with respect to permutations of the columns of the matrix $\J$
we can always place this minimum element in the position $(1,1)$.
Now let us redefine
the elements of the {\it first} line as follows:

\be
\lb{15}
J_{1j} = J^{(1)} + \tl J_{1j} \; , \; \; \; \; \; (j \not= 1)
\ee
and leave all the other elements unchanged. According to (\ref{3}),
the elements $\tl J_{1j}$ are described by {\it the same} 
exponential distribution: $P[\tl J_{1j}] = \exp\( -\tl J_{1j}\), \; 
(\tl J_{1j} \geq 0)$, while for $J^{(1)}$ the distribution
is: 

\be
\lb{16}
P[J^{(1)}] = N \exp\( - N J^{(1)} \)
\ee
Due to the constrains (\ref{1}) the above redefinition 
produces only simple shift of the Hamiltonian (\ref{2}):

\be
\lb{17}
H = J^{(1)} + \sum_{i,j=1}^{N} S_{ij} \tl \J
\ee
where, the random matrix $\tl \J$ contains "0" in the position (1,1),
while the rest of its elements are described by the same distribution (\ref{3}).
Now using the definition of $E_{N}$, eq.(\ref{5}), we can easily integrate 
out $J^{(1)}$ to get:

\be
\lb{18}
E_{N} = \fr{1}{N} + E_{N}^{(1)}
\ee
where

\be
\lb{19}
E_{N}^{(1)} \; \; = \; \; {\bf E} \( \; \; \;
\begin{tabular}{|l|l|l|l|l|l|}
\hline
0 & $\;$ & $\;$ & $\;$ & ... & $\;$ \\
\hline
 &  &  &  & ... & \\ 
\hline
 &  &  &  & ... & \\ 
\hline
 &  &  &  & ... & \\ 
\hline
\multicolumn{6}{|c|}{. . . . . . . . . . . . }  \\ 
\hline
 &  &  &  & ... & \\ 
\hline
\end{tabular}
\; \; \; \)
\ee

To calculate $E_{N}^{(1)}$ let us consider the {\it second} line of the 
above random matrix, and 
among $N$ its elements $J_{2j}$ let us find the minimum one:
$J^{(2)} \equiv \min_{j}(J_{2j})$. Now, due to "0" in the position (1,1)
the first column of this matrix is no more equivalent to the rest
of $(N-1)$ columns (which remain to be equivalent among themselves).
Therefore, with the probability $1/N$ the minimum element can be in the 
position (2,1), and with the probability $(N-1)/N$ it can be in 
the rest of the positions of the second line, and in this last case we can 
place it in the position (2,2). Then we shit the values of the 
elements of the second line: $J_{2j} = J^{(2)} + \tl J_{2j}$ 
(which leave the distribution of $\{\tl J_{2j}\}$ unchanged). 
The integration over $J^{(2)}$ gives one more factor $1/N$, and for 
$E_{N}$ we get:

\be
\lb{20}
E_{N} = \fr{2}{N} + \fr{(N-1)}{N} E_{N}^{(2)} + \fr{1}{N} \tl E_{N}^{(2)}
\ee
where

\be
\lb{21}
E_{N}^{(2)} \; \; = \; \; {\bf E} \( \; \; \;
\begin{tabular}{|l|l|l|l|l|l|}
\hline
0 &  & $\;$ & $\;$ & ... & $\;$ \\
\hline
 & 0 &  &  & ... & \\ 
\hline
 &  &  &  & ... & \\ 
\hline
 &  &  &  & ... & \\ 
\hline
\multicolumn{6}{|c|}{. . . . . . . . . . . . }  \\ 
\hline
 &  &  &  & ... & \\ 
\hline
\end{tabular}
\; \; \; \)
\ee
and 

\be
\lb{22}
\tl E_{N}^{(2)} \; \; = \; \; {\bf E} \( \; \; \;
\begin{tabular}{|l|l|l|l|l|l|}
\hline
0 & $\;$ & $\;$ & $\;$ & ... & $\;$ \\
\hline
0 &  &  &  & ... & \\ 
\hline
 &  &  &  & ... & \\ 
\hline
 &  &  &  & ... & \\ 
\hline
\multicolumn{6}{|c|}{. . . . . . . . . . . . }  \\ 
\hline
 &  &  &  & ... & \\ 
\hline
\end{tabular}
\; \; \; \)
\ee
Eq.(\ref{20}) can be represented in the form:

\be
\lb{23}
E_{N} = \fr{2}{N} + E_{N}^{(2)} + \fr{1}{N} \d E_{N}^{(2)}
\ee
where

\be
\lb{24}
\d E_{N}^{(2)} \; \; = \; \; {\bf E} \( \; \; \;
\begin{tabular}{|l|l|l|l|l|l|}
\hline
0 & $\;$ & $\;$ & $\;$ & ... & $\;$ \\
\hline
0 &  &  &  & ... & \\ 
\hline
 &  &  &  & ... & \\ 
\hline
 &  &  &  & ... & \\ 
\hline
\multicolumn{6}{|c|}{. . . . . . . . . . . . }  \\ 
\hline
 &  &  &  & ... & \\ 
\hline
\end{tabular}
\; \; \; \) \; \; - \; \;
{\bf E} \( \; \; \;
\begin{tabular}{|l|l|l|l|l|l|}
\hline
0 &  & $\;$ & $\;$ & ... & $\;$ \\
\hline
 & 0 &  &  & ... & \\ 
\hline
 &  &  &  & ... & \\ 
\hline
 &  &  &  & ... & \\ 
\hline
\multicolumn{6}{|c|}{. . . . . . . . . . . . }  \\ 
\hline
 &  &  &  & ... & \\ 
\hline
\end{tabular}
\; \; \; \)
\ee
To calculate the value $E_{N}^{(2)}$ defined by the matrix:

\be
\lb{25}
\begin{tabular}{|l|l|l|l|l|l|}
\hline
0 &  & $\;$ & $\;$ & ... & $\;$\\
\hline
 & 0 &  &  & ... & \\ 
\hline
 &  &  &  & ... & \\ 
\hline
 &  &  &  & ... & \\ 
\hline
\multicolumn{6}{|c|}{. . . . . . . . . . . . }  \\ 
\hline
 &  &  &  & ... & \\ 
\hline
\end{tabular}
\ee
let us consider its {\it third} line, and 
among $N$ elements $J_{3j}$ let us find the minimum one:
$J^{(3)} \equiv \min_{j}(J_{3j})$. Due to two "0" in the positions (1,1)
and (2,2)
the first and the second columns of this matrix are equivalent between themselves,
but they are not equivalent to the rest
of $(N-2)$ columns (which remain to be equivalent among themselves).
Therefore, with the probability $2/N$ the minimum element can be placed in the 
position (3,2), and with the probability $(N-2)/N$ it can be in 
the rest of the positions of the third line, and here we can 
place it in the position (3,3). Then we shit the values of the 
elements of the third line: $J_{3j} = J^{(3)} + \tl J_{3j}$ 
(which again leave the distribution of $\{\tl J_{3j}\}$ unchanged), and 
integrate over $J^{(3)}$ which gives one more factor $1/N$. In this way 
we get:

\be
\lb{26}
E_{N} = \fr{3}{N} + \fr{(N-2)}{N} E_{N}^{(3)} + \fr{2}{N} \tl E_{N}^{(3)}
+ \d E_{N}^{(2)}
\ee
where

\be
\lb{27}
E_{N}^{(3)} \; \; = \; \; {\bf E} \( \; \; \;
\begin{tabular}{|l|l|l|l|l|l|}
\hline
0 &  &  & $\;$ & ... & $\;$ \\
\hline
 & 0 &  &  & ... & \\ 
\hline
 &  & 0 &  & ... & \\ 
\hline
 &  &  &  & ... & \\ 
\hline
\multicolumn{6}{|c|}{. . . . . . . . . . . . }  \\ 
\hline
 &  &  &  & ... & \\ 
\hline
\end{tabular}
\; \; \; \)
\ee
and 

\be
\lb{28}
\tl E_{N}^{(3)} \; \; = \; \; {\bf E} \( \; \; \;
\begin{tabular}{|l|l|l|l|l|l|}
\hline
0 &  & $\;$ & $\;$  & ... & $\;$ \\
\hline
 & 0 &  &  & ... & \\ 
\hline
 & 0 &  &  & ... & \\ 
\hline
 &  &  &  & ... & \\ 
\hline
\multicolumn{6}{|c|}{. . . . . . . . . . . . }  \\ 
\hline
 &  &  &  & ... & \\ 
\hline
\end{tabular}
\; \; \; \)
\ee
Eq.(\ref{26}) can be represented in the form:

\be
\lb{29}
E_{N} = \fr{3}{N} + E_{N}^{(3)} + \fr{2}{N} \d E_{N}^{(3)} + 
\fr{1}{N} \d E_{N}^{(2)}
\ee
where

\be
\lb{30}
\d E_{N}^{(3)} \; \; = \; \; {\bf E} \( \; \; \;
\begin{tabular}{|l|l|l|l|l|l|}
\hline
0 &   & $\;$ & $\;$ & ... & $\;$ \\
\hline
  & 0 &      &      & ... &      \\ 
\hline
  & 0 &      &      & ... &      \\ 
\hline
  &   &      &      & ... &      \\ 
\hline
\multicolumn{6}{|c|}{. . . . . . . . . . . . }  \\ 
\hline
  &   &      &      & ... &      \\ 
\hline
\end{tabular}
\; \; \; \) \; \; - \; \;
{\bf E} \( \; \; \;
\begin{tabular}{|l|l|l|l|l|l|}
\hline
0 &   &   & $\;$ & ... & $\;$ \\
\hline
  & 0 &   &      & ... &      \\ 
\hline
  &   & 0 &      & ... &      \\ 
\hline
  &   &   &      & ... &      \\ 
\hline
\multicolumn{6}{|c|}{. . . . . . . . . . . . }  \\ 
\hline
  &   &   &      & ... &      \\ 
\hline
\end{tabular}
\; \; \; \)
\ee
Proceeding in this way up to the last line we eventually get:

\be
\lb{31}
E_{N} = 1 + \sum_{k=2}^{N}\fr{k-1}{N} \d E_{N}^{(k)} 
\ee
(note that $E_{N}^{(N)} \equiv 0$ since it is given by the matrix
with all zeros on the diagonal)
where

\be
\lb{32}
\d E_{N}^{(k)} = {\bf E} \( \; \; 
\begin{tabular}{|l|l|l|l|l|l||l|l|l|}
\hline
 0&   &...&   &   & $\;$ & $\;$  & ...  & $\;$  \\
\hline
  & 0 &...&   &   &   &   & ...  &   \\ 
\hline
\multicolumn{9}{|c|}{. . . . . . . . . . . . . . . . . . . .}  \\
\hline
  &   &...& 0 &   &   &   & ...  &   \\ 
\hline
  &   &...&   & 0 &   &   & ...  &   \\ 
\hline
  &   &...&   & 0 &   &   & ...  &   \\
\hline
\hline
  &  &... &   &   &   &   & ...  &   \\  
\hline
\multicolumn{9}{|c|}{. . . . . . . . . . . . . . . . . . . .}  \\ 
\hline
  &  &... &   &   &   &   & ...  &   \\ 
\hline
\end{tabular}
\; \;  \) -
{\bf E} \( \; \; 
\begin{tabular}{|l|l|l|l|l|l||l|l|l|}
\hline
 0&   &...&   &   & $\;$ & $\;$  & ...  & $\;$  \\
\hline
  & 0 &...&   &   &   &   & ...  &   \\ 
\hline
\multicolumn{9}{|c|}{. . . . . . . . . . . . . . . . . . . .}  \\
\hline
  &   &...& 0 &   &   &   & ...  &   \\ 
\hline
  &   &...&   & 0 &   &   & ...  &   \\ 
\hline
  &   &...&   &   & 0 &   & ...  &   \\
\hline
\hline
  &  &... &   &   &   &   & ...  &   \\  
\hline
\multicolumn{9}{|c|}{. . . . . . . . . . . . . . . . . . . .}  \\ 
\hline
  &  &... &   &   &   &   & ...  &   \\ 
\hline
\end{tabular}
\; \;  \)
\ee
Here the double lines mark the positions of the $k$-th column and
the $k$-th line.

It can be proved (see Appendix A) that the above value $\d E_{N}^{(k)}$
is given by the {\it rectangular} $N\times k$ random matrix problem:

\be
\lb{33}
\d E_{N}^{(k)} = {\bf E} \( \; \; 
\begin{tabular}{|l|l|l|l|l|l||l|l|l|}
\hline
 0&   &...&   &   & $\;$ & $\;$  & ...  & $\;$  \\
\hline
  & 0 &...&   &   &   &   & ...  &   \\ 
\hline
\multicolumn{9}{|c|}{. . . . . . . . . . . . . . . . . . . .}  \\
\hline
  &   &...& 0 &   &   &   & ...  &   \\ 
\hline
  &   &...&   & 0 &   &   & ...  &   \\ 
\hline
  &   &...&   & 0 &   &   & ...  &   \\
\hline
\hline
\end{tabular}
\; \;  \)
\ee
defined by the Hamiltonian:

\be
\lb{34}
H[\hat S;\hat J] = \sum_{i=1}^{N}\sum_{j=1}^{k} S_{ij} \J
\ee
where the random matrix $\J$ is shown in eq.(\ref{33}) (with the same
independent exponential distributions of non-zero elements).
Here the "truncated"  $N\times k$ part of the original permutation
matrix $\hat S$ again can have only one "1" in each line, and besides
it has $k$ columns each containing
only one "1" and $(N-k)$ columns each containing only "0".

It turns out that the above "rectangular" problem, eq.(\ref{33})
can be solved explicitly (the proof see in Appendix B):

\be
\lb{37}
\d E_{N}^{(k)} = \fr{1}{k(k-1)}\sum_{l=1}^{k-1} \fr{l}{N-l}
\ee
Substituting this result into eq.(\ref{31}) we find:

\be
\lb{38}
E_{N} = 1 + \fr{1}{N}\sum_{k=2}^{N}\fr{1}{k} \sum_{l=1}^{k-1} \fr{l}{N-l}  
\ee
After simple algebra one eventually finds:

\be
\lb{39}
E_{N} - E_{N-1} = \fr{1}{N^{2}}
\ee
which proves the result (\ref{13}).

\vspace{5mm}

It should be noted in conclusion that obtained solution is valid only
for the considered exponential type distribution, eq.(\ref{3}). It is
crucial for the above proof that the form of the distribution 
of a random element $\J$ doesn't change after its shift on a constant
value. On the other hand, it is clear from the above proof
that in the thermodynamic limit $N\to\infty$ the leading (in $1/N$)
contribution to $E_{N}$ is defined only by the very beginning
of the distribution, $P[J\to 0]$. Therefore, the result
$E_{N\to\infty} = \z(2)$ must be correct, also for the "rectangular" type
distribution: $P[0\leq J\leq 1] = 1; \; P[J>1] = 0$ (it is actually the 
model with this type of distribution which was studied in the 
replica solution \cite{MP}). For the discussion of other types of the
matching models see e.g. \cite{martin} and references therein.

\sectio{Appendix A}

In this Appendix we prove that value of $\d E_{N}^{(k)}$
defined in eq.(\ref{32}) 
is given by the rectangular $N\times k$ problem (\ref{33}).

First, let us consider the most simple case $k=2$: 

\be
\lb{40}
\d E_{N}^{(2)} = {\bf E} \( \; \; 
\begin{tabular}{|l||l|l|l|l|l|l|l|}
\hline
 0& $\;$ & $\;$ & $\;$ & $\;$ & $\;$ & ... & $\;$  \\
\hline
 0&      &      &      &      &      & ... &       \\ 
\hline
\hline
  &      &      &      &      &      & ... &       \\ 
\hline
  &      &      &      &      &      & ... &       \\ 
\hline
  &      &      &      &      &      & ... &       \\
\hline
  &      &      &      &      &      & ... &       \\  
\hline
\multicolumn{8}{|c|}{. . . . . . . . . . . . . . . . . }  \\ 
\hline
  &      &      &      &      &      & ... &       \\ 
\hline
\end{tabular}
\; \;  \) -
{\bf E} \( \; \; 
\begin{tabular}{|l|l||l|l|l|l|l|l|}
\hline
 0&      & $\;$ & $\;$ & $\;$ &  $\;$ & ... & $\;$  \\
\hline
  &   0  &      &      &      &       & ... &       \\ 
\hline
\hline
  &      &      &      &      &       & ... &       \\ 
\hline
  &      &      &      &      &       & ... &       \\ 
\hline
  &      &      &      &      &       & ... &       \\
\hline
  &      &      &      &      &       & ... &       \\  
\hline
\multicolumn{8}{|c|}{. . . . . . . . . . . . . . . . .}  \\ 
\hline
  &      &      &      &      &       & ... &       \\ 
\hline
\end{tabular}
\; \;  \) \; \equiv \tl E_{N}^{(2)} - E_{N}^{(2)}
\ee
The above two problems, $\tl E_{N}^{(2)}$ and $E_{N}^{(2)}$,
differ only by the permutation of two elements: (2,1) and (2,2),
while all the other matrix elements $\J$ in both matrices are
the same.
Nevertheless, even this "tiny" permutation, in general,
can make the ground state configurations of the matrix $\hat S$
in the two problems to be quite different.
Note that for the calculation of the above average
energy difference $\d E_{N}^{(2)}$ we can make averaging over 
$\J$ both simultaneously (keeping $\J$ to be the same in both
problems) as well as separately for $\tl E_{N}^{(2)}$ and for
$E_{N}^{(2)}$.

For further proof it is important to introduce the concept
of {\it equivalence} among the columns (and among the lines).
We call the two
columns $j_1$ and $j_2$ (or the two lines $i_1$ and $i_2$) 
equivalent if the probabilities of the positions
$(i,j_1)$ and $(i,j_2)$ (or $(i_1,j)$ and $(i_2,j)$) to be occupied 
in the ground state are equal.

Due to obvious symmetry properties of the systems under consideration,
it is evident that in each of the 
above problems, $\tl E_{N}^{(2)}$ and $E_{N}^{(2)}$,
all the columns on the right of the double vertical line,
and all the lines below the double horizontal line are
equivalent among themselves. 
On the other hand, the first two lines in each of the above problems
are also equivalent between themselves, but they are not equivalent
to the rest of the $(N-2)$ lines. Besides, in the problem $E_{N}^{(2)}$
we have first two columns which are equivalent between themselves,
but which are not equivalent to the rest of the $(N-2)$ columns.
Finally in the problem $\tl E_{N}^{(2)}$ the first column is not 
equivalent to the rest of the $(N-1)$ columns.

The energy difference $\d E_{N}^{(2)}$, eq(\ref{40}), can be 
represented as follows:

\be
\lb{41}
\d E_{N}^{(2)} = \ol{\(\sum_{j=1}^{N} \[J_{\tl i(j)j} - J_{i(j)j}\]\)}
\ee
where $J_{\tl i(j)j}$ and $J_{i(j)j}$ represent the elements of the
$j$-the column  occupied in the 
ground states of the problems $\tl E_{N}^{(2)}$ and $E_{N}^{(2)}$ 
correspondingly. 

In any ground state in each of these problems
there are two columns in which the occupied elements belongs 
to one the first two lines. Therefore, in the summation over the 
columns in eq.(\ref{41}) we can find two, three or four columns
in which one of the elements (or both) of the difference
$(J_{\tl i(j)j} - J_{i(j)j})$ belongs to one of the first two
lines. In the rest of the columns (their number can be $(N-2)$, $(N-3)$
or $(N-4)$) both the element $(\tl i(j),j)$ and the element $(i(j),j)$
belong to the rest of $(N-2)$ {\it equivalent} lines. Now,
if with some probability we have the ground states in which 
the element $(\tl i(j),j)$ is occupied in the problem
$\tl E_{N}^{(2)}$  and the element $(i(j),j)$ is occupied in the problem
$E_{N}^{(2)}$, then {\it with the same probability} we must have 
the inverse situation: the element $(\tl i(j),j)$ is occupied in the problem
$E_{N}^{(2)}$, while the element $(i(j),j)$ is occupied in the problem
$\tl E_{N}^{(2)}$. Since both problems are defined by {\it the same}
matrix $\J$ (except for the two elements (2,1) and (2,2)) in the averaging
over $\J$ in eq.(\ref{41}) the contributions from all the columns with
no elements in one of the first two lines must cancel.

Thus, to compute the difference $\d E_{N}^{(2)}$, we should take care
only of those columns which in the ground states of the problems
$\tl E_{N}^{(2)}$ and $E_{N}^{(2)}$ contain elements of the first two lines.

Due to the equivalence of the first two lines
and due to equivalence of the $(N-2)$ columns ($j=3,...N)$ 
we can reduce all the relevant
ground states of the problem $E_{N}^{(2)}$ to the following four
non-equivalent basic configurations:

\be
\lb{42}
\begin{tabular}{|l|l||l|l|l}
\hline
 0&   & $\;\;$   & $\bullet$ & $\;\;$ \\
\hline
  & 0 & $\bullet$&           &        \\ 
\hline
\hline
  &   &          &           &        \\
\multicolumn{5}{c}{ (a) }             \\ 
\end{tabular}
\; ; \; \;
\begin{tabular}{|l|l||l|l|l}
\hline
 0& $\bullet$  & $\;\;$   & $\;\;$ & $\;\;$ \\
\hline
  & 0          & $\bullet$&        &        \\ 
\hline
\hline
  &            &          &        &        \\
\multicolumn{5}{c}{ (b) }             \\ 
\end{tabular}
\; ; \; \; 
\begin{tabular}{|l|l||l|l|l}
\hline
$\bigodot$ &            & $\;\;$   & $\;\;$ & $\;\;$ \\
\hline
           & 0          & $\bullet$&        &      \\ 
\hline
\hline
  &            &      &           &      \\
\multicolumn{5}{c}{ (c) }             \\ 
\end{tabular}
\; ; \; \; 
\begin{tabular}{|l|l||l|l|l}
\hline
$\bigodot$&           & $\;\;$ & $\;\;$    & $\;\;$ \\
\hline
           &$\bigodot$&        &           &        \\ 
\hline
\hline
          &           &        &           &        \\
\multicolumn{5}{c}{ (d) }             \\ 
\end{tabular}
\ee
where "$\bullet$" represent the elements occupied in the ground state
configuration of the matrix $\hat S$, and "$\bigodot$" denote 
occupied element "0". Note that each of the above configurations
represents the whole set of equivalent configurations. 
For instance, (\ref{42}(b)) represents all configurations with
"$\bullet$" in any of $(N-2)$ positions $(2,j)$, $(j=3,...,N)$,
as well as all configurations with "$\bullet$" in the position (2,1)
and another "$\bullet$" in any of $(N-2)$ positions $(1,j)$, $(j=3,...,N)$.
The diagram (\ref{42}(c)) represents all configurations in which
{\it one} of the zeros is occupied.
Note also that the configurations of the type:

\be
\lb{43} 
\begin{tabular}{|l|l||l|l|l}
\hline
 0         & $\bullet$  & $\;\;$   & $\;\;$ & $\;\;$ \\
\hline
$\bullet$  & 0          &          &        &        \\ 
\hline
\hline
  &            &          &        &        \\ 
\end{tabular}
\ee
must be excluded from the consideration since they {\it can not} be
the ground state being always higher in energy than the states
represented in (\ref{42}(d)).

Due to equivalence of $(N-1)$ columns $(j=2,...,N)$ in
the problem $\tl E_{N}^{(2)}$ here we have only two
non-equivalent basic configurations:

\be
\lb{44} 
\begin{tabular}{|l||l|l|l|l}
\hline
 0& $\ast$  & $\;\;$   & $\;\;$ & $\;\;$ \\
\hline
 0&            & $\ast$&        &        \\ 
\hline
\hline
  &            &          &        &        \\
\multicolumn{5}{c}{ (a) }             \\  
\end{tabular}
\; \; ; \; \; \; \; \; \; \; \;
\begin{tabular}{|l||l|l|l|l}
\hline
$\bigodot$ &           & $\;\;$ & $\;\;$ & $\;\;$ \\
\hline
 0         & $\ast$ &        &        &        \\ 
\hline
\hline
  &            &      &           &      \\ 
\multicolumn{5}{c}{ (b) }             \\ 
\end{tabular}
\ee
Here for the occupied positions we use the notation "$\ast$" 
instead of "$\bullet$" to distinguish them from the ones
in the ground states of the problem $E_{N}^{(2)}$.

Now to compute the contribution to the difference of the energies
$\d E_{N}^{(2)}$, eq.(\ref{41}), we have to consider all possible 
combinations of the ground state configurations of the problem
$E_{N}^{(2)}$, eq.(\ref{42}), and of the ones of the problem
$\tl E_{N}^{(2)}$, eq.(\ref{44}). 

It is evident that if in the problem $E_{N}^{(2)}$ we have one of 
the configurations of the type (\ref{42}(a)) or (\ref{42}(b)) and in the
problem $\tl E_{N}^{(2)}$ we have one of the configurations of the
type (\ref{44}(a)) (all those in which no one "0" is occupied),
then (since the two problems contain the same set of $\J$'s)
the positions of "$\bullet$" and "$\ast$" must coincide.
Therefore, these two cases give no contribution to 
$\d E_{N}^{(2)}$, eq.(\ref{41}).

It is also evident that the combination of one of the ground states
of the type (\ref{42}(a)) or (\ref{42}(b)) with (\ref{44}(b)) is
impossible. For example, let us suppose that the ground state of the
problem $E_{N}^{(2)}$ is the configuration (\ref{42}(a)), and the
one of the problem $\tl E_{N}^{(2)}$ is the configuration (\ref{44}(b)).
Then, according to the definition of the ground state, the energy
of (\ref{42}(a)) must be smaller than the one of the configuration
(\ref{42}(d)), which in turn (since the problem $\tl E_{N}^{(2)}$
contain the same set of $\J$'s) must be smaller than the energy
of the configuration (\ref{44}(b)). On the other hand, the energy
of the configuration (\ref{42}(a)) is equal to the one 

\be
\lb{45} 
\begin{tabular}{|l||l|l|l|l}
\hline
 0& $\;\;$   & $\;\;$ & $\ast$ & $\;\;$ \\
\hline
 0&          & $\ast$ &        &        \\ 
\hline
\hline
  &            &          &        &        \\
\end{tabular}
\ee
of the problem $\tl E_{N}^{(2)}$. Thus, the energy of (\ref{45})
is smaller than the one of (\ref{44}(b)), and therefore (\ref{44}(b))
can not be the ground state.

Similar arguments shows that the combinations of 
(\ref{42}(c)) with (\ref{44}(a)), as well as
(\ref{42}(d)) with (\ref{44}(a))  
are also impossible.

The combination of (\ref{42}(c)) and (\ref{44}(b)) is allowed, but in 
this case, according to the definition of the ground state, 
the position of "$\bullet$" in (\ref{42}(c)) of 
the problem $E_{N}^{(2)}$ must coincide with the position of "$\ast$" 
in (\ref{44}(b)) of the problem $\tl E_{N}^{(2)}$, and therefore
this combination also gives no contribution to $\d E_{N}^{(2)}$.

Finally we are left with the combination of the ground state
configurations of the types (\ref{42}(d)) and (\ref{44}(b)) 
which indeed give finite contribution to $\d E_{N}^{(2)}$, eq.(\ref{41}).
According to the discussion below eq.(\ref{41}), here we get no contribution
from all the elements $\J$ of all $(N-2)$ lines $(i=3,...N)$,
and thus all these elements fall out of the computation.
The difference of energies between (\ref{44}(b)) and (\ref{42}(d))
which does enter into the computation
is equal to one of $2(N-1)$ elements of the first two lines
of the problem $\tl E_{N}^{(2)}$ (note again, that the diagram (\ref{44}(b))
represents one of $2(N-1)$ configurations with one occupied "0").
Therefore, the average difference $\d E_{N}^{(2)}$ can be represented 
in the form of the "truncated" problem:

\be
\lb{46}
\d E_{N}^{(2)} = {\bf E} \( \; \; 
\begin{tabular}{|l||l|l|l|l|l|l|l|}
\hline
 0& $\;$ & $\;$ & $\;$ & $\;$ & $\;$ & ... & $\;$  \\
\hline
 0&      &      &      &      &      & ... &       \\ 
\hline
\end{tabular} \; \;  \)
\ee
where the ground state is just one the configurations 
of the type (\ref{44}(b)).

The generalization of the proof for arbitrary $k$ is straightforward.
First of all, it is evident that to get the proof for a general
value of $k$ it is actually sufficient to consider the 
problem with $k=3$:

\be
\lb{47}
\d E_{N}^{(3)} = {\bf E} \( \; \; 
\begin{tabular}{|l|l||l|l|l|l|l|l|}
\hline
 0& $\;$ & $\;$ & $\;$ & $\;$ & $\;$ & ... & $\;$  \\
\hline
  &  0   &      &      &      &      & ... &       \\ 
\hline
  &  0   &      &      &      &      & ... &       \\ 
\hline
\hline
  &      &      &      &      &      & ... &       \\ 
\hline
  &      &      &      &      &      & ... &       \\
\hline
  &      &      &      &      &      & ... &       \\  
\hline
\multicolumn{8}{|c|}{. . . . . . . . . . . . . . . . . }  \\ 
\hline
  &      &      &      &      &      & ... &       \\ 
\hline
\end{tabular}
\; \;  \) -
{\bf E} \( \; \; 
\begin{tabular}{|l|l|l||l|l|l|l|l|}
\hline
 0&      & $\;$ & $\;$ & $\;$ &  $\;$ & ... & $\;$  \\
\hline
  &   0  &      &      &      &       & ... &       \\ 
\hline
  &      &  0   &      &      &       & ... &       \\ 
\hline
\hline
  &      &      &      &      &       & ... &       \\ 
\hline
  &      &      &      &      &       & ... &       \\
\hline
  &      &      &      &      &       & ... &       \\  
\hline
\multicolumn{8}{|c|}{. . . . . . . . . . . . . . . . .}  \\ 
\hline
  &      &      &      &      &       & ... &       \\ 
\hline
\end{tabular}
\; \;  \) \; \equiv \tl E_{N}^{(3)} - E_{N}^{(3)}
\ee
because here the first line and the first columns can "represent"
all the other equivalent lines and columns of the problems with
$k > 3$.

Then we should make quite similar to eqs.(\ref{42}) and (\ref{44})
classification (which turns out to be only slightly more cumbersome)
of all non-equivalent ground state configurations of the problems
$E_{N}^{(3)}$ and $\tl E_{N}^{(3)}$ according to the positions
of the occupied elements of the first three lines.
Simple analysis shows that here
again the only relevant (for $\d E_{N}^{(3)}$) configurations 
of the problem $E_{N}^{(3)}$ are the ones with all three zeros occupied,
while in the problem $\tl E_{N}^{(3)}$ these are the configurations
with one or two of the zeros occupied. On the other hand, due to 
equivalence of the rest of $(N-3)$ lines $(i=4,...N)$ one finds that
all the elements $\J$ of these lines fall out of the computation.
In this way one obtains that energy difference $\d E_{N}^{(3)}$
is given by two or one of the elements of the three first lines
of the problem $\tl E_{N}^{(3)}$, which is just defined in terms
of the "truncated" problem:

\be
\lb{48}
\d E_{N}^{(3)} = {\bf E} \( \; \; 
\begin{tabular}{|l|l||l|l|l|l|l|l|}
\hline
 0& $\;$ & $\;$ & $\;$ & $\;$ & $\;$ & ... & $\;$  \\
\hline
  &  0   &      &      &      &      & ... &       \\ 
\hline
  &  0   &      &      &      &      & ... &       \\ 
\hline
\end{tabular}
\; \;  \)
\ee

The calculation of the actual values of $\d E_{N}^{(k)}$
is presented in the next Appendix.

\sectio{Appendix B}

In this Appendix we prove that

\be
\lb{49}
\d E_{N}^{(k)} = \fr{1}{k(k-1)}\sum_{l=1}^{k-1} \fr{l}{N-l}
\ee
The solution of the case $k=2$, eq.(\ref{46}), is trivial.
Here the ground state configuration is of the type (\ref{44}(b)),
where the position of "$\ast$" must be at the smallest element
out of $2(N-1)$ non-zero $\J$'s. According to the distribution
(\ref{3}), for the average value of this element we get:

\be
\lb{50}
\d E_{N}^{(2)} = \fr{1}{2(N-1)}
\ee

Now let us consider slightly more complicated case $k=3$, eq.(\ref{48}).
Simple analysis of the the structures of possible ground state configurations
shows that all of them can be taken into account in terms of only
$3\times 3$ matrix:

\be
\lb{51}
\begin{tabular}{|l|l|l|}
\hline
 0         & $\otimes$ & z         \\
\hline
 $\otimes$ &      0    & $\otimes$ \\ 
\hline
 y         &      0    & x         \\ 
\hline
\end{tabular}
\ee
Here $x$ is the smallest element out of $2(N-2)$ equivalent elements 
$J_{2j}$ $(j=3,...,N)$ and $J_{3j}$ $(j=3,...,N)$ of the second 
and the third lines;
$y$ is the smallest element out of two equivalent elements $J_{21}$ and $J_{31}$;
$z$ is the smallest element out of $(N-2)$ equivalent 
elements $J_{1j}$ $(j=3,...,N)$ of the first line; 
the symbol "$\otimes$" denotes the elements which do not enter into
any ground state configuration. One can easily check that the matrix
in eq.(\ref{48}) can have only two ground state energies equal to $x$ or
equal to $(y+z)$. Note that if we consider this problem in terms of the
$3\times 3$ matrix (\ref{51}), the element
$x$ could as well be placed in  
the position (3,2) (instead of "$\otimes$" which then should be placed at
the position (3,3)), as well as $y$ could be interchanged with "$\otimes$"
in the positions (2,1) and (3,1). 

Now one can easily note that original $3\times 3$ problem  (\ref{51})
is actually equivalent to the $2\times 2$ problem:

\be
\lb{52}
\d E_{N}^{(3)} = {\bf E} \( \; \; 
\begin{tabular}{|l|l|}
\hline
 0         &  z         \\
\hline
 y         &  x         \\ 
\hline
\end{tabular}
\; \;  \)
\ee
where, according to the definitions of the random parameters $x, y$ and $z$
their statistical distributions are:

\be
\lb{53}
P(x) = 2(N-2) \exp\[-2(N-2) x\]
\ee

\be
\lb{54}
P(y) = 2 \exp\(-2 y \)
\ee

\be
\lb{55}
P(z) = (N-2) \exp\[-(N-2) z\]
\ee
Keeping in mind further generalization of the solution for an arbitrary $k$,
we solve the problem (\ref{52}) in the following way. 
Similarly to the procedure described in the beginning of Section 2,
we can "shift" the elements of the second line ($x$ and $y$)
by the value of the smallest of them, and then integrate it out:

\be
\lb{56}
\d E_{N}^{(3)} = \fr{1}{2(N-1)} + \fr{1}{(N-1)} 
{\bf E} \( \; \; 
\begin{tabular}{|l|l|}
\hline
 0         &  z         \\
\hline
 0         &  x         \\ 
\hline
\end{tabular}
\; \;  \)
\ee 
The factor $1/(N-1)$ in the second term of the above equation
is the probability that $y$ is smaller than $x$ (if the smallest
element is $x$, then the remaining problem will have all zeros at the 
diagonal, and the minimum energy of this problem is identically equal to zero).
The solution of the remaining $2\times 2$ problem is trivial, and eventually
we get the following result:

\be
\lb{57}
\d E_{N}^{(3)} = \fr{1}{2(N-1)} +  \fr{1}{3(N-1)(N-2)} =
                 \fr{1}{3\cdot 2}\[\fr{1}{N-1} + \fr{2}{N-2} \] 
\ee

Now the generalization of the above procedure 
for an arbitrary value of $k$ becomes evident.
First we note that all possible ground state configurations of the
$N\times k$ problem $\d E_{N}^{(k)}$, eq.(\ref{33}), can be 
taken into account in terms of the $k\times k$
matrix: 

\be
\lb{58}
\begin{tabular}{|l|l|l|l|l|l|}
\hline
 0        &           &...&            & $\otimes$ & $z_1$      \\
\hline
          & 0         &...&            & $\otimes$ & $z_2$      \\ 
\hline
\multicolumn{6}{|c|}{. . . . . . . . . . . . . . . . . . . . .}               \\
\hline
          &           &...& 0          & $\otimes$ & $z_{(k-2)}$\\ 
\hline
$\otimes$ & $\otimes$ &...& $\otimes$  & 0         & $\otimes$  \\ 
\hline
 $y_1$    & $y_2$     &...& $y_{(k-2)}$& 0         &  $x$       \\
\hline
\end{tabular}
\ee
Here $x$ is the smallest element out of $2(N-k+1)$ equivalent elements 
$J_{(k-1)j}$ $(j=k,...,N)$ and $J_{kj}$ $(j=k,...,N)$ of the last two lines; 
$y_j$ $(j=1,...,(k-2)$ is the smallest element out of two equivalent 
elements $J_{(k-1)j}$ and $J_{kj}$;
$z_i$ $(i=1,...,(k-2)$ is the smallest element out of $(N-k+1)$ equivalent 
elements $J_{ij}$ $(j=k,...,N)$ of the $i$-th line; and again
the symbol "$\otimes$" denotes the elements which do not enter into
any ground state configuration.

According to the above definitions of the random parameters
$x, \{y_j\}$ and $\{z_i\}$ their probability distribution functions
are:

\be
\lb{59}
P(x) = 2(N-k+1) \exp\[-2(N-k+1) x\]
\ee

\be
\lb{60}
P(y_j) = 2 \exp\(-2 y_j \)
\ee

\be
\lb{61}
P(z_i) = (N-k+1) \exp\[-(N-k+1) z_i\]
\ee
In this way we can reduce the calculation of $\d E_{N}^{(k)}$ to the
$(k-1)\times (k-1)$ matrix problem:

\be
\lb{62}
\d E_{N}^{(k)} = {\bf E} \( \; \; 
\begin{tabular}{|l|l|l|l|l|}
\hline
 0        &           &...&            & $z_1$      \\
\hline
          & 0         &...&            & $z_2$      \\ 
\hline
\multicolumn{5}{|c|}{. . . . . . . . . . . . . . . . . }          \\
\hline
          &           &...& 0          & $z_{(k-2)}$\\ 
\hline
 $y_1$    & $y_2$     &...& $y_{(k-2)}$& $x$       \\
\hline
\end{tabular}
\; \;  \)
\ee
Taking into account the equivalence of the first $(k-2)$ columns
here we can integrate out the smallest element of the last line to get:

\be
\lb{63}
\d E_{N}^{(k)} = \fr{1}{2(N-1)} + \fr{k-2}{(N-1)} 
{\bf E} \( \; \; 
\begin{tabular}{|l|l|l|l|l|l|}
\hline
 0        &           &...&            &       & $z_1$      \\
\hline
          & 0         &...&            &       & $z_2$      \\ 
\hline
\multicolumn{6}{|c|}{. . . . . . . . . . . . . . . . . }          \\
\hline
          &           &...&  0         &       & $z_{(k-3)}$ \\
\hline
          &           &...&            & 0     & $z_{(k-2)}$\\ 
\hline
 $y_1$    & $y_2$     &...& $y_{(k-3)}$& 0     & $x$       \\
\hline
\end{tabular}
\; \;  \)
\ee
Now one can easily see that all possible ground state
configurations in the remaining $(k-1)\times (k-1)$ problem 
can be taken into account 
in the same way as in the previous $k\times k$ one, eq.(\ref{58}).
Here we can 
reduce the number of relevant elements by choosing the smallest
one between $x$ and $z_{(k-2)}$, as well as between each $y_j$
of the last line and $J_{(k-2)j}$ $(j=1,...,(k-3))$ of the previous line.
In this way we get:

\be
\lb{64}
\d E_{N}^{(k)} = \fr{1}{2(N-1)} + \fr{k-2}{(N-1)} 
{\bf E} \( \; \; 
\begin{tabular}{|l|l|l|l|l|l|}
\hline
 0        &           &...&            &  $\otimes$ & $z_1$      \\
\hline
          & 0         &...&            &  $\otimes$ & $z_2$      \\ 
\hline
\multicolumn{6}{|c|}{. . . . . . . . . . . . . . . . . }          \\
\hline
          &           &...&  0         & $\otimes$  & $z_{(k-3)}$ \\
\hline
$\otimes$ & $\otimes$ &...& $\otimes$  & 0          & $\otimes$ \\ 
\hline
 $y_1$    & $y_2$     &...& $y_{(k-3)}$& 0          & $x$       \\
\hline
\end{tabular}
\; \;  \)
\ee
where the random elements $x$ and  $\{y_j\}$ according to their
definitions are now described by the following  
distribution functions:

\be
\lb{65}
P(x) = 3(N-k+1) \exp\[-3(N-k+1) x\]
\ee

\be
\lb{66}
P(y_j) = 3 \exp\(-3 y_j \)
\ee
while the distribution functions of $z_i$'s remain unchanged, eq.(\ref{61}).
In this way we can reduce the calculation of $\d E_{N}^{(k)}$ to the
$(k-2)\times (k-2)$ matrix problem:

\be
\lb{67}
\d E_{N}^{(k)} = \fr{1}{2(N-1)} + \fr{k-2}{(N-1)} 
{\bf E} \( \; \; 
\begin{tabular}{|l|l|l|l|l|}
\hline
 0        &           &...&            &  $z_1$      \\
\hline
          & 0         &...&            &  $z_2$      \\ 
\hline
\multicolumn{5}{|c|}{. . . . . . . . . . . . . . . . . }          \\
\hline
          &           &...&  0         & $z_{(k-3)}$ \\
\hline
 $y_1$    & $y_2$     &...& $y_{(k-3)}$& $x$       \\
\hline
\end{tabular}
\; \;  \)
\ee
Here again we can integrate out the smallest element of the last line
to get:

\be
\lb{68}
\d E_{N}^{(k)} = \fr{1}{2(N-1)} + \fr{k-2}{(N-1)} \[
\fr{1}{3(N-2)} + \fr{k-3}{(N-2)}
{\bf E} \( \; \; 
\begin{tabular}{|l|l|l|l|l|l|}
\hline
 0        &           &...&            &       & $z_1$      \\
\hline
          & 0         &...&            &       & $z_2$      \\ 
\hline
\multicolumn{6}{|c|}{. . . . . . . . . . . . . . . . . }          \\
\hline
          &           &...&  0         &       & $z_{(k-4)}$ \\
\hline
          &           &...&            & 0     & $z_{(k-3)}$\\ 
\hline
 $y_1$    & $y_2$     &...& $y_{(k-4)}$& 0     & $x$       \\
\hline
\end{tabular}
\; \;  \) \]
\ee
Continuing these iterations till the last trivial $2\times 2$ problem
we eventually obtain the following result:

\be
\lb{69}
\d E_{N}^{(k)} = 
\fr{1}{2(N-1)} + \fr{k-2}{(N-1)} \[
\fr{1}{3(N-2)} + \fr{k-3}{(N-2)} \[ 
\fr{1}{4(N-3)} + \fr{k-4}{(N-3)} \[...\[
\fr{1}{k(N-k+1)} \]...\]\]\]
\ee
After simple algebra the above expression can be easily reduced to
the following form: 

\be
\lb{70}
\d E_{N}^{(k)} = \fr{1}{k(k-1)} \[
\fr{1}{N-1} + \fr{2}{N-2} + ... + \fr{k-1}{N-k+1} \]
\ee
which proves eq.(\ref{49}).

\vspace{15mm}

{\bf Acknowledgements}.
The author is grateful to M.M\'ezard, S.Bravy and G.Parisi 
for numerous useful discussions.

\end{document}